\documentclass{article}
\usepackage{arxiv}
\usepackage{graphicx}%
\usepackage{multirow}%
\usepackage{amsmath,amssymb,amsfonts}%
\usepackage{amsthm}%
\usepackage{mathrsfs}%
\usepackage[title]{appendix}%
\usepackage{xcolor}%
\usepackage{textcomp}%
\usepackage{manyfoot}%
\usepackage{booktabs}%
\usepackage{algorithm}%
\usepackage{algorithmicx}%
\usepackage{algpseudocode}%
\usepackage{listings}%
\usepackage{authblk}
\usepackage[pdfpagelabels,hidelinks]{hyperref}   

\theoremstyle{thmstyleone}%
\theoremstyle{thmstyletwo}%
\theoremstyle{thmstylethree}%
\title{Global quantum phase estimation via hybrid quantum--classical learning}

\author[1,2]{Qingchuan Yang}
\author[1]{Xianing Feng}
\author[1]{Lianfu Wei}

\affil[1]{Information Quantum Technology Laboratory, School of Information Science and Technology, Southwest Jiaotong University, Chengdu 610031, Sichuan, China}
\affil[2]{Sichuan Zhitu Linghai Intelligent Technology Co., Ltd., Chengdu 610213, China}

\date{\today}

\begin{document}
\maketitle

\footnotetext[1]{These authors contributed equally to this work. Emails: yangqingchuan@my.swjtu.edu.cn, fengxn5@swjtu.edu.cn}
\footnotetext[2]{Corresponding author. Email: lfwei@swjtu.edu.cn}

\begin{abstract}
Achieving both high precision and large dynamic range remains a central challenge in quantum metrology, as improving local sensitivity typically reduces the unambiguous estimation range. Variational quantum interferometers enhance precision but are generally limited to narrow operating regimes. Here we introduce a hybrid variational quantum--classical neural network interferometer (VQ-CNNI), where a shallow quantum circuit encodes phase-dependent measurement statistics and a neural network performs nonlinear phase reconstruction. Joint optimization enables accurate and unambiguous phase estimation over $[-\pi,\pi)$ without loss of precision. We show that this performance requires co-optimization of quantum encoding and classical decoding. Visualization of the learned representation geometry links global estimation to well-conditioned measurement statistics across the full phase range, enabling stable inversion. Odd-symmetric activations further improve robustness by promoting global consistency. These results suggest that global quantum metrology can be understood through the learnability of the quantum--classical representation, providing a practical route to programmable interferometers with both high precision and large dynamic range.
\end{abstract}


\section{Introduction}

Quantum sensing exploits superposition, entanglement, and coherence to achieve measurement sensitivities beyond classical limits, enabling applications such as optical atomic clocks, gravitational-wave detection, biomedical imaging, and precision navigation~\cite{ludlow2015optical,tse2019quantum,taylor2016quantum,chen2019single}. Many of these applications reduce to phase estimation, where physical quantities including electromagnetic fields, inertial signals, and gravitational perturbations are encoded as phase shifts in interferometric systems~\cite{abbott2016observation,degen2017quantum}.

A major challenge in quantum metrology is the trade-off between estimation precision and dynamic range. Classical protocols are limited by the standard quantum limit (SQL), where the estimation error scales as $1/\sqrt{N}$ with the number of probe particles $N$. By contrast, entanglement-enhanced protocols can in principle achieve Heisenberg-limited scaling proportional to $1/N$~\cite{giovannetti2004quantum,davis2016approaching,yin2023experimental}. However, improved local sensitivity is typically accompanied by a substantial reduction in the range of unambiguous estimation. For example, highly entangled probe states such as GHZ and NOON states achieve optimal precision only within a narrow interval $[-\pi/N,\pi/N]$, outside of which phase ambiguity emerges~\cite{higgins2009demonstrating,gorecki2022multiple,gorecki2020pi}. This trade-off persists even in protocols that surpass the SQL via combined entanglement and squeezing~\cite{feng2024beating}. This limitation is particularly problematic for sensing applications that require both high precision and wide dynamic range, including quantum gyroscopes, radio-frequency sensing, and long-baseline interferometry~\cite{herbschleb2021ultra,luo2024high}.

Numerous approaches have been proposed to mitigate the trade-off between sensitivity and dynamic range. Iterative and adaptive protocols can expand the unambiguous estimation range, but often require repeated coherent evolution, adaptive feedback, and long coherence times~\cite{liu2023full}. Alternative strategies based on geometric phases, nonlinear encoding or quantum deamplification can partially extend the operational range, typically at the cost of increased circuit complexity or reduced sensitivity~\cite{Arai2018geometric,liu2025enhancing}. Variational quantum interferometers (VQIs) provide a programmable framework for optimizing probe preparation and measurement~\cite{kaubruegger2021quantum,marciniak2022optimal}, yet improving global performance generally demands deeper circuits and more sophisticated control, limiting robustness on noisy intermediate-scale quantum devices~\cite{qi2024variational,dong2025optimal}. In parallel, hybrid quantum--classical learning architectures have emerged as promising tools for near-term quantum technologies. Parameterized quantum circuits combined with classical neural networks can be trained end-to-end on measurement outcomes, exhibiting strong capability in quantum feature extraction and supervised learning~\cite{benedetti2019parameterized,cerezo2021variational,schuld2018supervised,abbas2021power,schuld2020circuit,perez2020data}. Related machine-learning methods have also been explored in quantum metrology for calibration, adaptive control, and sensing optimization~\cite{cimini2019calibration,nolan2021machine,gebhart2023learning,fiderer2021neural,maclellan2024end}. These developments suggest the possibility of achieving globally accurate phase estimation through joint quantum--classical learning rather than increasing quantum circuit complexity alone. However, whether such hybrid architectures can overcome the precision--range trade-off using shallow quantum resources remains largely unexplored.

Here we introduce a hybrid variational quantum--classical neural network interferometer (VQ-CNNI), in which a shallow variational quantum circuit generates phase-dependent measurement statistics and a neural network learns the inverse mapping from measurement outcomes to phase over the full interval $\phi\in[-\pi,\pi)$. We show that globally accurate phase estimation can be achieved without increasing quantum circuit depth through end-to-end joint optimization of quantum encoding and classical decoding. Under finite-shot measurements, VQ-CNNI substantially outperforms a shallow VQI baseline with a single encoding and decoding layer (enc=dec=1), whereas decoupled two-stage optimization fails to reproduce this advantage. We further show that accurate global estimation emerges when the learned measurement distribution forms a smooth and globally distinguishable quantum feature embedding that supports stable inversion by the decoder. These results identify representation learnability and geometric conditioning as central principles for global quantum metrology, establishing joint quantum--classical optimization as a mechanism for organizing quantum measurement representations to enable robust global phase estimation.

\section{Methodology}
\label{sec:Methodology}

\subsection{Hybrid quantum--classical framework for global phase estimation}

Quantum phase estimation determines the parameter $\phi$ encoded by the unitary evolution $U_{\phi}=e^{-i\phi J_z}$. Variational quantum interferometers (VQIs) optimize probe preparation and decoding through parameterized unitaries $U_{\mathrm{En}}(\boldsymbol{\theta})$ and $U_{\mathrm{De}}(\boldsymbol{\vartheta})$, producing the conditional probability distribution
\begin{equation}
p_{\boldsymbol{\theta},\boldsymbol{\vartheta}}(m|\phi)
=\left|\langle m|U_{\mathrm{De}}(\boldsymbol{\vartheta})U_{\phi}U_{\mathrm{En}}(\boldsymbol{\theta})|\downarrow\rangle^{\otimes N}\right|^2,
\end{equation}
from which an estimator $\phi_{\mathrm{est}}(m)$ is constructed~\cite{yang2020probe}. Conventional VQIs are typically trained by minimizing the Bayesian mean squared error (BMSE) under a narrow prior, while performance is benchmarked against the quantum Fisher information (QFI) and the quantum Cram\'er--Rao bound (QCRB), which define the ultimate local precision limit~\cite{marciniak2022optimal}. These objectives are well suited for local estimation because they characterize infinitesimal distinguishability around a fixed operating point. Global phase estimation over $\phi\in[-\pi,\pi)$, however, additionally requires that distinct phases generate distinguishable measurement distributions and that the inverse mapping from measurement statistics to phase remain well conditioned across the full periodic domain.

Extending local optimization strategies to larger estimation ranges generally requires deeper circuits or more sophisticated control, which remain challenging on noisy intermediate-scale quantum devices~\cite{kaubruegger2021quantum}. Rather than increasing quantum circuit complexity, we investigate whether globally accurate phase estimation can emerge from fixed shallow quantum resources through joint quantum--classical optimization. To isolate this effect, we restrict both the conventional VQI baseline and our hybrid model to the same minimal quantum architecture with a single encoding and decoding layer (enc=dec=1), ensuring identical quantum expressivity. We therefore introduce the variational quantum--classical neural network interferometer (VQ-CNNI), in which a classical neural network performs nonlinear inversion from quantum-generated measurement features.

\begin{figure}[t]
\centering
\includegraphics[width=0.95\textwidth]{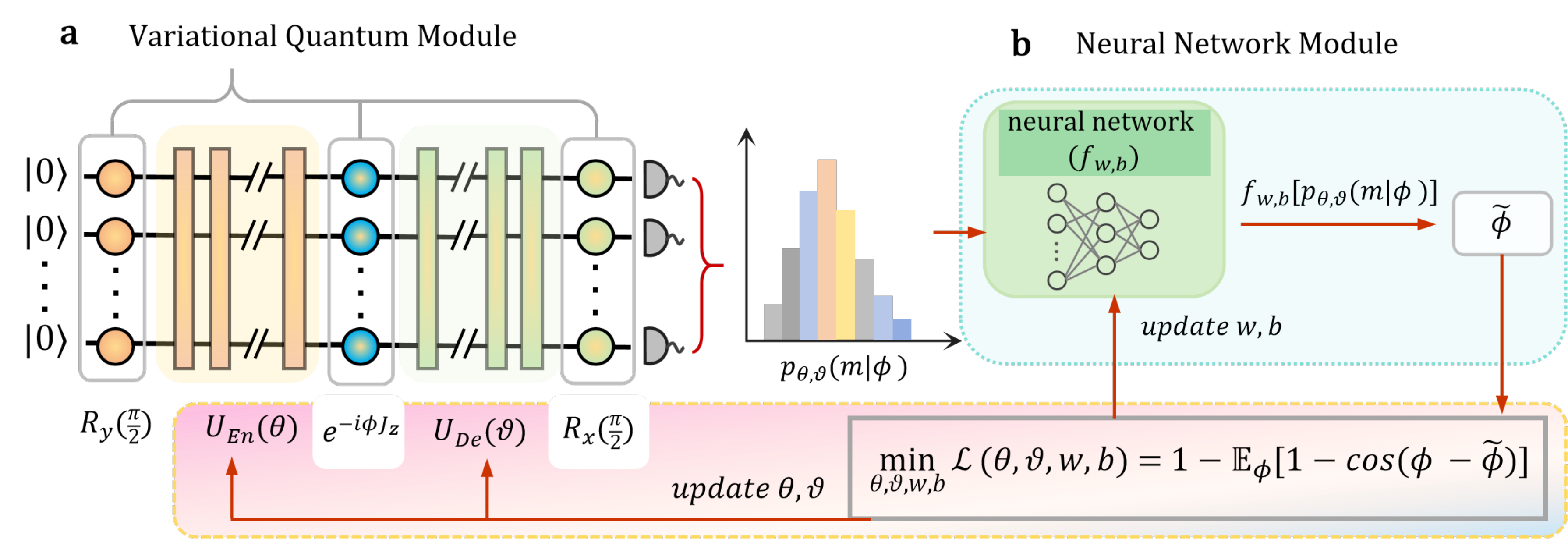}
\caption{
\textbf{Joint quantum--classical framework for global phase estimation.}
\textbf{a} Variational quantum module: a shallow interferometric circuit produces the measurement distribution $p_{\boldsymbol{\theta},\boldsymbol{\vartheta}}(m|\phi)$.
\textbf{b} Classical inference module: a neural network outputs the phase estimate $\tilde{\phi}$ using a circular representation. All parameters are jointly optimized over the full periodic domain.
}
\label{fig:framework}
\end{figure}

The VQ-CNNI pipeline $\phi \rightarrow p_{\boldsymbol{\theta},\boldsymbol{\vartheta}}(m|\phi) \rightarrow \tilde{\phi}$ integrates a parametrized quantum circuit with a classical neural network (parameters $\mathbf{w},\mathbf{b}$), jointly optimized as $\boldsymbol{\Theta}=\{\boldsymbol{\theta},\boldsymbol{\vartheta},\mathbf{w},\mathbf{b}\}$ (Fig.~\ref{fig:framework}). The quantum module follows a Ramsey-type interferometer:
\begin{equation}
U_R(\phi;\boldsymbol{\theta},\boldsymbol{\vartheta})
=
R_x\!\left(\frac{\pi}{2}\right)
U_{\mathrm{De}}(\boldsymbol{\vartheta})
R_z(\phi)
U_{\mathrm{En}}(\boldsymbol{\theta})
R_y\!\left(\frac{\pi}{2}\right),
\end{equation}
where $U_{\mathrm{En}}$ and $U_{\mathrm{De}}$ consist of collective rotations $\mathcal{R}_{\alpha}(\beta)=e^{-i\beta J_{\alpha}}$ and twisting operations $\mathcal{T}_{\alpha}(\chi)=e^{-i\chi J_{\alpha}^{2}}$. Projective measurements yield the probabilities
\begin{equation}
p_{\boldsymbol{\theta},\boldsymbol{\vartheta}}(m|\phi) = \left|\langle m|U_R(\phi;\boldsymbol{\theta},\boldsymbol{\vartheta})|0\rangle^{\otimes N}\right|^2,
\end{equation}
where $m\in\{-N/2,\ldots,N/2\}$ labels the population imbalance between spin states~\cite{kaubruegger2021quantum}.

To account for phase periodicity, the neural network predicts the circular representation $(\sin\tilde{\phi},\cos\tilde{\phi})$, from which the phase estimate is recovered as $\tilde{\phi}=\operatorname{atan2}(\sin\tilde{\phi},\cos\tilde{\phi})$. All parameters are trained using the circular loss
\begin{equation}
\mathcal{L}(\boldsymbol{\Theta})
= 1 - \mathbb{E}_{\phi\sim\mathcal{U}([-\pi,\pi))}
\left[\cos(\phi-\tilde{\phi})\right],
\end{equation}
evaluated using quadrature over $N_{\phi}$ uniformly sampled phases in $[-\pi,\pi)$ (we use $N_{\phi}=100$ during training)~\cite{lecamwasam2021investigations}. Unlike BMSE optimization under a narrow Gaussian prior, this objective enforces consistent estimation performance across the entire $2\pi$ interval. The quantum circuit architecture is matched to the shallow VQI baseline (enc=dec=1), ensuring that performance improvements arise from the joint optimization strategy and nonlinear classical neural network decoder (decoder) rather than increased quantum expressivity. Estimation performance is evaluated using both finite-shot sampling ($10^6$ shots) and exact expectation values. The VQI baseline is trained using BMSE minimization~\cite{marciniak2022optimal} for comparison. Algorithm~\ref{alg:vqcnni} in Appendix~\ref{appendix:alg:vq-cnni} summarizes the complete joint optimization procedure.

\subsection{Characterizing global phase estimation}

To characterize global phase estimation, we introduce two quantities complementing local Fisher information: the decoding Jacobian, which measures the conditioning of the learned inverse mapping, and the squared wrapped phase error, which quantifies periodic estimation accuracy.

\subsubsection{Decoding Jacobian analysis}
To characterize the conditioning of the learned inverse mapping (that is, whether small changes in the input produce correspondingly small changes in the output), we analyze the derivative of the estimated phase with respect to the true phase,
\begin{equation}
J(\phi) = \frac{d\tilde{\phi}}{d\phi},
\end{equation}
which we refer to as the decoding Jacobian. This quantity measures the local gain of the decoder: for an ideal estimator satisfying $\tilde{\phi}\approx\phi$, the Jacobian should remain close to unity across the entire periodic interval. We characterize the inverse mapping using both the mean and variance of $J(\phi)$:
\begin{equation}
\bar{J} = \mathbb{E}_{\phi}[J(\phi)],
\qquad
\sigma_J^2 = \mathrm{Var}(J(\phi)).
\end{equation}
The mean $\bar{J}$ quantifies global calibration: $\bar{J}<1$ indicates systematic compression of phase variations, whereas $\bar{J}>1$ indicates amplification. The variance $\sigma_J^2$ measures fluctuations in local sensitivity. Small $\sigma_J^2$ indicates that the decoder maintains nearly uniform sensitivity across the phase domain, whereas large fluctuations reveal locally compressed or stretched regions that degrade inversion stability.

\subsubsection{Performance metrics}

To quantify estimation accuracy for periodic variables, we employ the squared wrapped phase error (SWPE). Given a true phase $\phi$ and its estimate $\tilde{\phi}$, the wrapped phase error is defined as
\begin{equation}
\delta\phi = \mathrm{Arg}\left(e^{i(\tilde{\phi}-\phi)}\right),
\qquad
\delta\phi\in[-\pi,\pi),
\end{equation}
where $\mathrm{Arg}(\cdot)$ denotes the principal argument. The squared wrapped phase error is then expressed in decibel scale as
\begin{equation}
\mathrm{SWPE(dB)} = 10\log_{10}(\delta\phi)^2.
\end{equation}
This metric naturally respects the $2\pi$ periodicity of the phase variable and provides an intuitive measure of global estimation fidelity.

\section{Results and Discussion}
\label{sec:results}

The shallow VQI baseline (enc=dec=1) achieves high local precision near a fixed operating point, but fails to maintain accurate estimation across the full phase interval $\phi\in[-\pi,\pi)$. By contrast, the VQ-CNNI learns a globally invertible mapping between measurement statistics and phase. We evaluate performance using the squared wrapped phase error (SWPE) and analyze the learned representations through quantum feature heatmaps and PCA visualizations.

\subsection{Global phase estimation beyond the local regime}
\label{subsec:global_learnability}
\begin{figure}[t]
\centering
\includegraphics[width=0.85\textwidth]{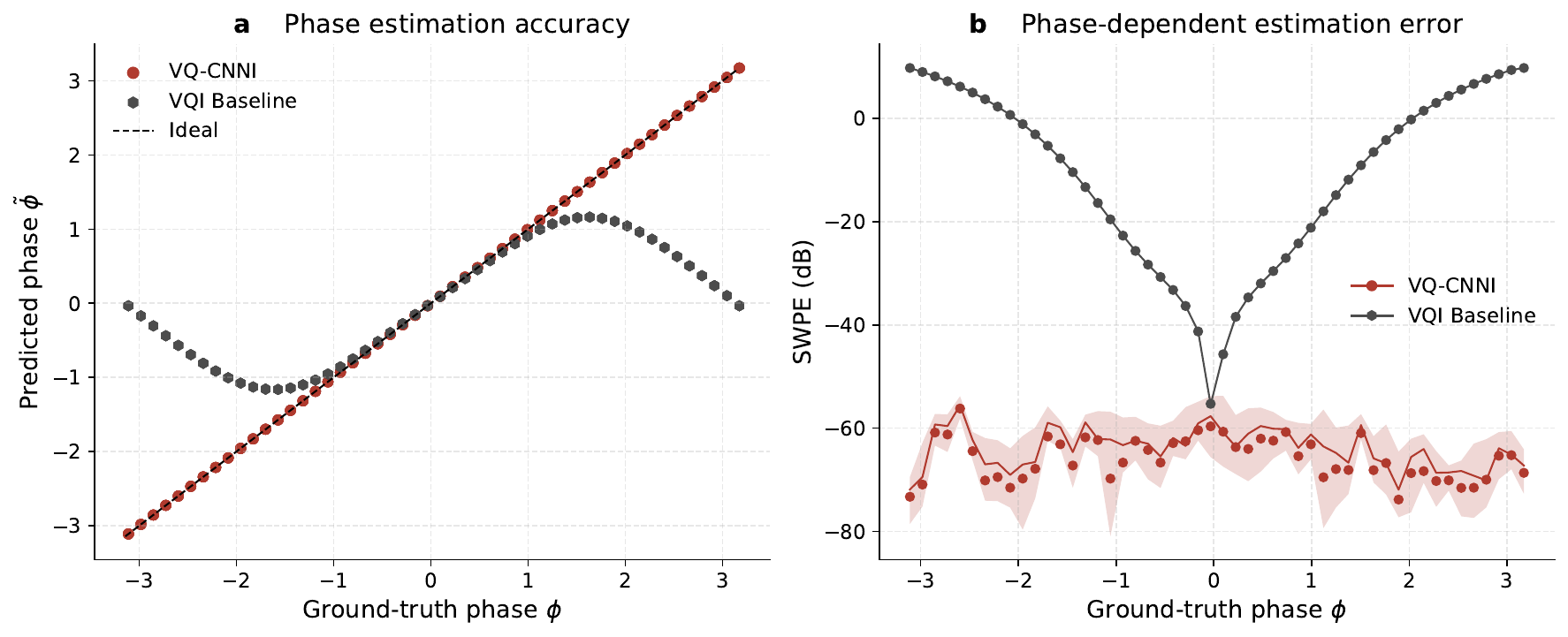}
\caption{
\textbf{Global phase estimation and error distribution.}
\textbf{a} Predicted phase $\tilde{\phi}$ versus ground-truth phase $\phi$ across 20 independent runs. Markers indicate mean predictions, and the dashed line denotes the ideal relation $\tilde{\phi}=\phi$. VQI baseline exhibits strong deviations and branch folding away from its operating point, whereas VQ-CNNI preserves a monotonic global mapping.
\textbf{b} Squared wrapped phase error (SWPE) as a function of $\phi$. Solid lines show the median across runs, shaded regions indicate interquartile ranges, and markers denote mean values. VQ-CNNI maintains uniformly low error across the full phase domain.
}
\label{fig:perf}
\end{figure}
Fig.~\ref{fig:perf} reveals the distinction between the shallow VQI baseline and the Softsign-based VQ-CNNI across the full phase range. As shown in Fig.~\ref{fig:perf}a, the VQI baseline estimates phase accurately only near its operating point, while predictions away from this region exhibit strong nonlinearity and branch folding, where distinct phases map to similar outputs. This behavior reflects limited global distinguishability. By contrast, VQ-CNNI preserves a monotonic mapping across the entire periodic domain, with predictions remaining close to the ideal relation $\tilde{\phi}=\phi$. The corresponding SWPE distributions are shown in Fig.~\ref{fig:perf}b. The VQI baseline exhibits low error only within a narrow local region, whereas VQ-CNNI maintains uniformly low error across the full $2\pi$ interval, demonstrating stable global phase estimation. All results in Fig.~\ref{fig:perf} are evaluated under finite-shot measurements ($10^6$ shots), explicitly incorporating sampling noise. To analyze the learned representations underlying this performance difference, the following sections evaluate expectation values directly, thereby eliminating measurement shot noise.

\subsection{Joint optimization shapes quantum feature and latent geometries}

Global phase estimation requires that distinct phases produce distinguishable measurement statistics across the full periodic domain. In our framework, the measurement distribution $p_{\boldsymbol{\theta},\boldsymbol{\vartheta}}(m|\phi)$ acts as a learned representation of phase, referred to as the quantum feature embedding. Its direct visualization is denoted the quantum feature heatmap, while its PCA projection is referred to as the quantum feature manifold. The neural network decoder further transforms this representation into an internal latent representation, whose PCA projection is referred to as the latent manifold.

\begin{figure}[t]
\centering
\includegraphics[width=0.95\textwidth]{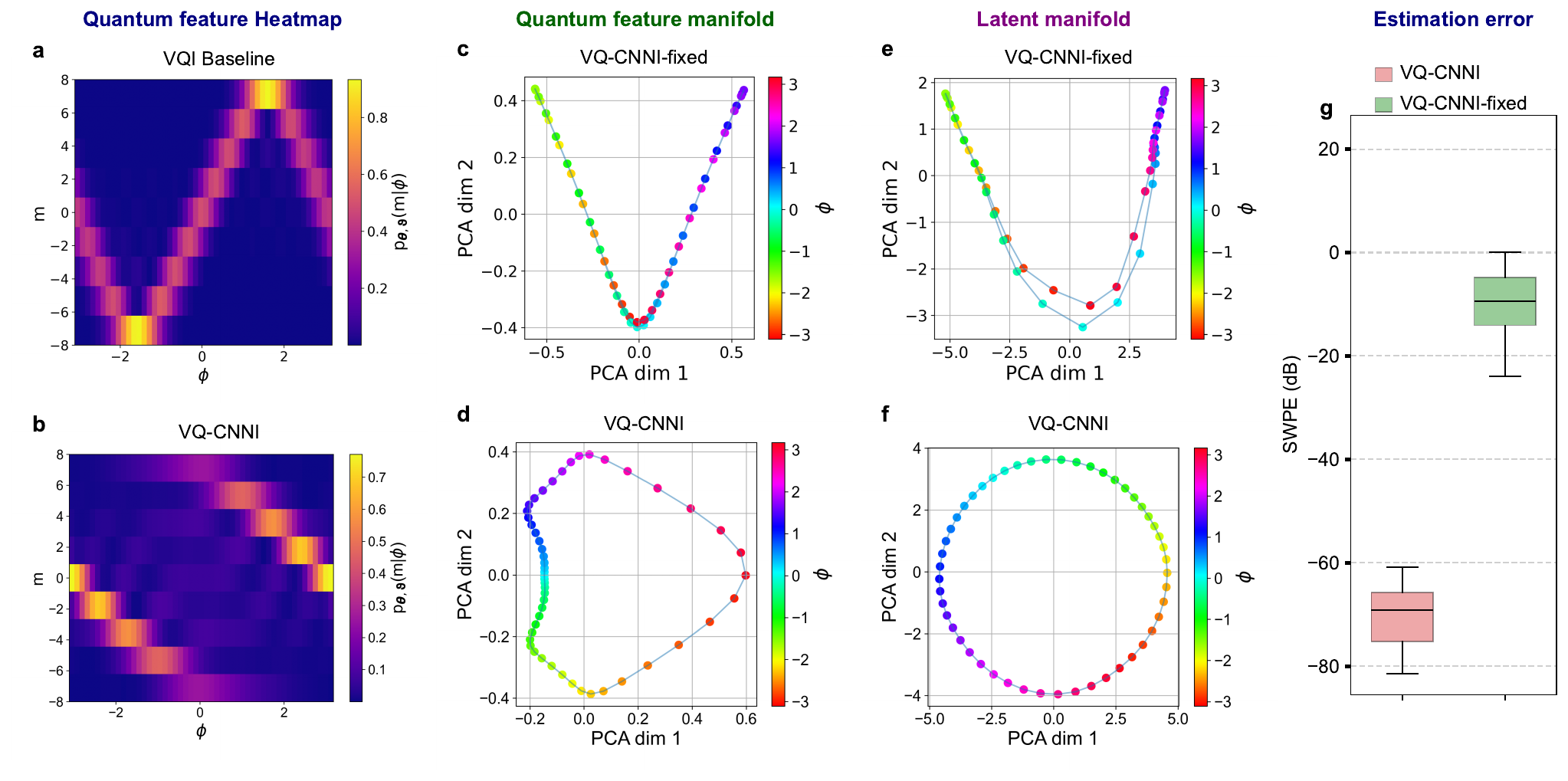}
\caption{
\textbf{Quantum feature embeddings and latent manifolds under different optimization strategies.}
\textbf{a,b} Quantum feature heatmaps for the VQI baseline (\textbf{a}) and jointly optimized VQ-CNNI (\textbf{b}). VQI exhibits oscillatory repeating patterns, whereas VQ-CNNI produces an ordered phase evolution with strongly suppressed overlap between distant phases.
\textbf{c,d} PCA visualizations of the quantum feature manifolds for VQ-CNNI-fixed (\textbf{c}) and jointly optimized VQ-CNNI (\textbf{d}).
\textbf{e,f} Corresponding latent manifolds (PCA projections of the decoder's penultimate layer features) for VQ-CNNI-fixed (\textbf{e}) and jointly optimized VQ-CNNI (\textbf{f}).
\textbf{g} SWPE distributions across all test phases. 
}
\label{fig:distinguish}
\end{figure}

Figure~\ref{fig:distinguish} reveals the distinction between locally optimized variational interferometry and globally learnable phase estimation. The shallow VQI baseline trained using BMSE minimization produces oscillatory and repeating measurement patterns (Fig.~\ref{fig:distinguish}a), such that distant phases frequently generate similar measurement statistics. As a result, the corresponding quantum feature manifold lacks global organization and does not support stable inversion across the full periodic domain. By contrast, the jointly optimized VQ-CNNI produces an ordered quantum feature embedding with strongly suppressed phase overlap (Fig.~\ref{fig:distinguish}b). The corresponding PCA visualization reveals a globally organized manifold that preserves the periodic structure of the phase variable (Fig.~\ref{fig:distinguish}d). This organization is similarly reflected in the latent manifold learned by the decoder (Fig.~\ref{fig:distinguish}f), resulting in uniformly low SWPE across the full $2\pi$ interval (Fig.~\ref{fig:distinguish}g). To determine whether a classical neural network alone can recover globally accurate estimation from a locally optimized quantum feature embedding, we freeze the quantum parameters obtained from the VQI baseline (denoted VQ-CNNI-fixed) and train only the decoder. Although the nonlinear decoder partially reshapes the latent representation, the resulting quantum feature and latent manifolds remain geometrically distorted (Fig.~\ref{fig:distinguish}c,e), leading to substantially degraded global accuracy (Fig.~\ref{fig:distinguish}g).

These results show that globally accurate phase estimation requires joint optimization of the quantum embedding and decoder inverse mapping. The improvement does not arise solely from the classical decoder; rather, joint optimization reorganizes the geometry of the quantum measurement statistics itself, transforming locally ambiguous measurement statistics into a globally distinguishable and invertible representation suitable for robust global phase estimation.

\subsection{Two-stage joint optimization dynamics}

\begin{figure}[t]
\centering
\includegraphics[width=0.95\textwidth]{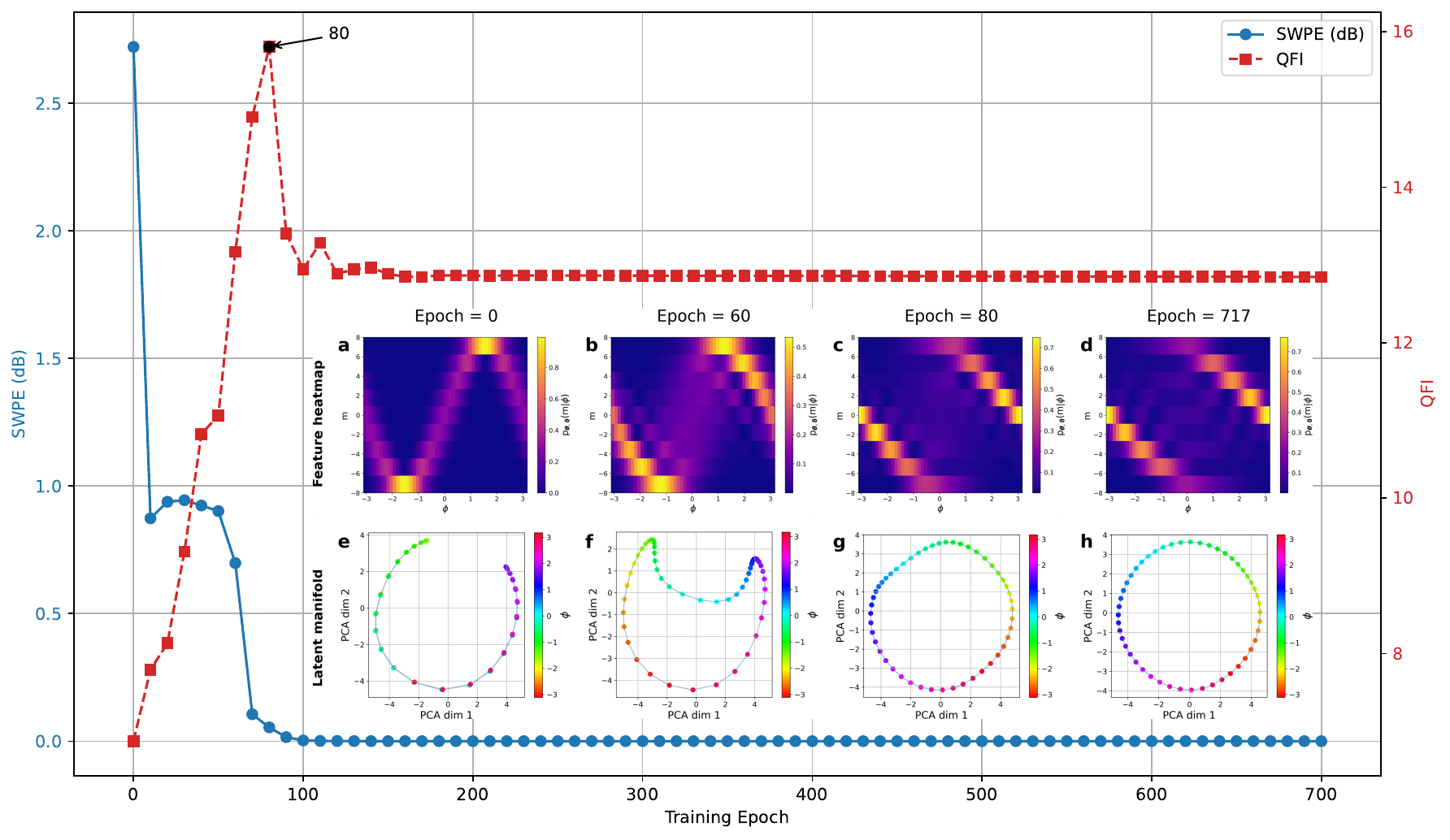}
\caption{
\textbf{Visualization of the joint optimization process.}
Evolution of QFI (red dashed) and SWPE (blue solid) during training. Insets show probability heatmaps (top) and latent manifolds (bottom) at selected epochs. The early stage establishes phase distinguishability and periodic topology, whereas the later stage primarily refines latent manifold geometry.
}
\label{fig:train}
\end{figure}

To understand how globally distinguishable representations emerge during training, we visualize the evolution of the quantum feature embedding, latent manifold, QFI, and SWPE throughout joint optimization. Figure~\ref{fig:train} reveals a clear two-stage optimization trajectory.
In the early stage (epochs 0--80), the model establishes phase distinguishability and periodic topology. At initialization, the quantum feature heatmap is highly degenerate and the latent manifold forms an open arc that does not encode phase periodicity. As training progresses, the measurement distributions become increasingly distinguishable, and the latent manifold closes into a complete, although irregular, circle. By epoch 80, the QFI reaches its maximum and phase degeneracy is largely eliminated. In the later stage (epochs 80--final), the quantum feature heatmap changes only weakly, whereas the latent manifold becomes progressively smoother and more uniformly organized. During this stage, the QFI marginally decreases and then saturates,  while the SWPE continues to decrease, indicating continued improvement of the global inverse mapping despite minimal change in local sensitivity.

These results reveal that joint optimization first establishes globally distinguishable quantum representations and then refines the geometry of the learned latent manifold to improve inverse-mapping stability. The continued reduction of SWPE after QFI saturation further demonstrates that robust global phase estimation depends not only on local sensitivity, but also on the global organization and conditioning of the learned representation.

\subsection{Odd symmetry enhances robustness of global estimation}
\label{subsec:robust}

Having established that global estimation performance is governed by representation geometry rather than local sensitivity alone, we next examine whether this advantage is specific to a particular activation function or reflects a broader principle of well-conditioned embeddings. We analyze several standard nonlinear activation functions, including Tanh, Arctan, Sigmoid, ELU, and Softsign. To isolate the role of symmetry, we further introduce a Softsign-shift activation,
\begin{equation}
\mathrm{Softsign\text{-}shift}(x)=\frac{x}{1+|x|}+1,
\end{equation}
which preserves the overall functional form while explicitly breaking odd symmetry.

\begin{figure}[htbp]
\centering
\includegraphics[width=0.95\textwidth]{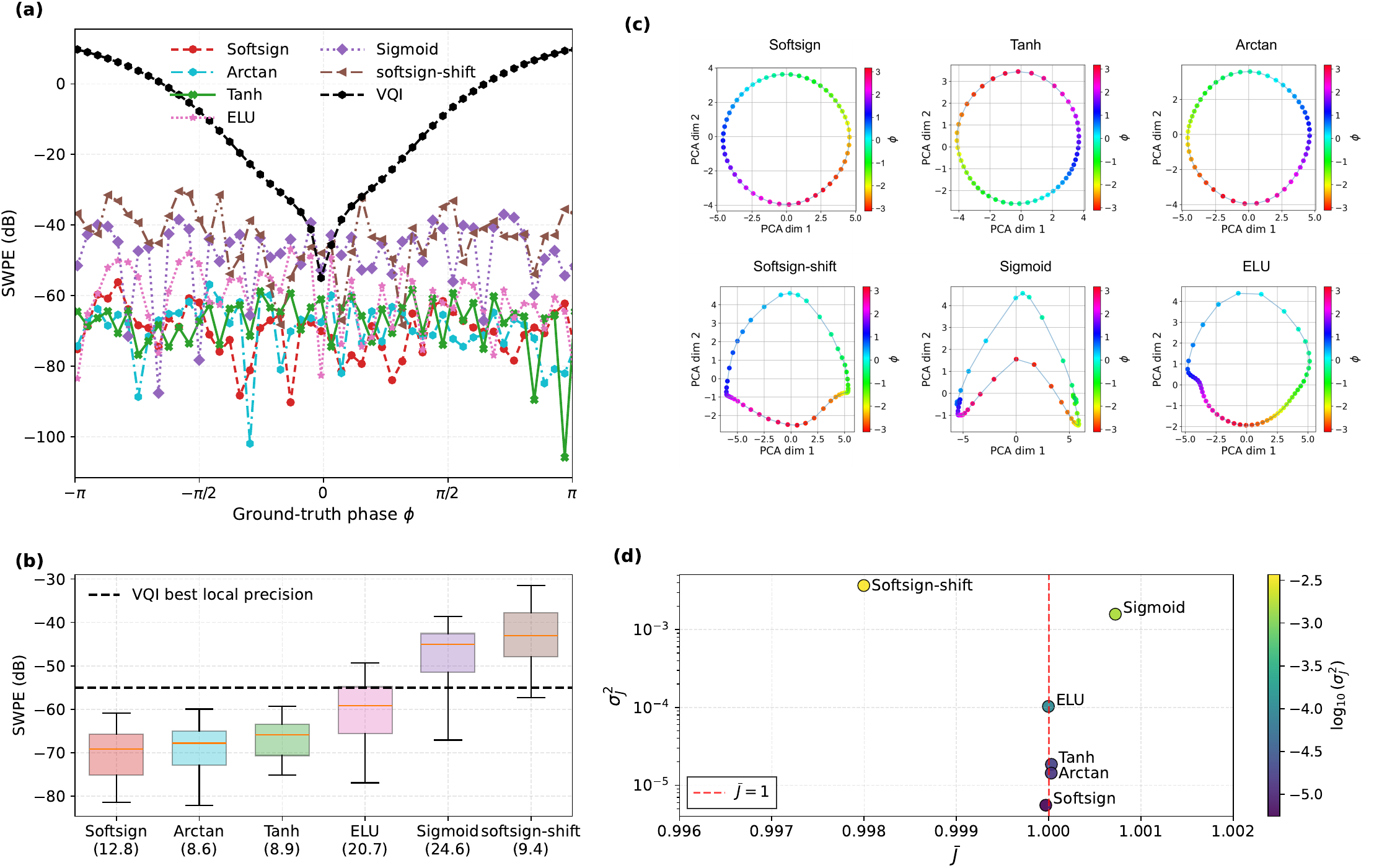}
\caption{
\textbf{Odd symmetry enhances robustness of global estimation.}
\textbf{a} SWPE (dB) as a function of the ground-truth phase $\phi$ for the VQI baseline and six VQ-CNNI variants with different activation functions.
\textbf{b} Boxplot comparison of SWPE distributions aggregated across all test phases for each model. The central line denotes the median, the box boundaries indicate the interquartile range (IQR), and the whiskers extend to the 5th and 95th percentiles. The dashed horizontal line marks the best local precision achieved by VQI. Numbers in parentheses below each model label indicate the corresponding QFI.
\textbf{c} Latent manifolds for different activation functions. Odd-symmetric activations preserve globally ordered and nearly circular manifolds aligned with the periodic structure of the phase variable, whereas asymmetric activations generate distorted manifolds with nonuniform parameterization.
\textbf{d} Decoding Jacobian statistics. Odd-symmetric activations cluster near $(\bar{J}=1,\sigma_J^2\to0)$, indicating globally uniform inverse mappings.
}
\label{fig:active}
\end{figure}

Figure~\ref{fig:active} shows that all activation functions achieve global phase estimation in the sense that predictions span the full interval $[-\pi,\pi)$ without catastrophic branch folding. The robustness and precision of the learned inverse mapping, however, vary substantially across activations. Odd-symmetric activations (Softsign, Tanh, and Arctan) maintain uniformly low SWPE across the full phase domain (Fig.~\ref{fig:active}a,b), whereas asymmetric activations such as Sigmoid, ELU, and Softsign-shift exhibit substantially larger estimation errors. These performance differences are directly reflected in the latent manifolds learned by the decoder (Fig.~\ref{fig:active}c). Odd-symmetric activations produce globally ordered and uniformly organized manifolds aligned with the periodic structure of the phase variable, while asymmetric activations generate distorted latent manifolds with nonuniform parameterization. The comparison between Softsign and Softsign-shift is particularly informative because it isolates the role of symmetry while preserving nearly the same nonlinear profile. Breaking odd symmetry degrades the median SWPE from approximately $-69.2\,\mathrm{dB}$ to $-43.0\,\mathrm{dB}$ (Fig.~\ref{fig:active}b), indicating that odd symmetry acts as an effective inductive bias for stabilizing the learned representation.

To quantify this effect, we evaluate the decoding Jacobian statistics introduced in Section~\ref{sec:Methodology}. As shown in Fig.~\ref{fig:active}d, odd-symmetric activations consistently satisfy the near-ideal condition $(\bar{J}\approx1,\sigma_J^2\rightarrow0)$, indicating well-conditioned inverse mappings. Softsign achieves the lowest Jacobian variance ($5.55\times10^{-6}$), followed by Arctan ($1.44\times10^{-5}$) and Tanh ($1.86\times10^{-5}$). By contrast, asymmetric activations display substantially larger fluctuations (ELU: $1.03\times10^{-4}$, Sigmoid: $1.58\times10^{-3}$, Softsign-shift: $3.68\times10^{-3}$), which directly correspond to degraded global estimation accuracy. Importantly, global estimation performance does not directly follow QFI ranking. Sigmoid and ELU exhibit comparatively high QFI values but inferior global estimation performance, whereas Softsign, Arctan, and Tanh achieve substantially lower SWPE despite lower QFI. These results indicate that robust global phase estimation depends not only on local sensitivity, but also on the geometric conditioning of the learned representation. Within this framework, odd symmetry acts as an effective geometric regularizer that stabilizes the latent manifold and improves global decoding robustness.

\section{Conclusion}

Achieving both high precision and large dynamic range remains a central challenge in quantum metrology because enhanced local sensitivity is often accompanied by a reduced range of unambiguous estimation. Here we introduced a hybrid variational quantum--classical neural network interferometer (VQ-CNNI), in which a shallow variational quantum circuit generates phase-dependent measurement statistics and a neural network performs nonlinear phase inversion. Through end-to-end joint optimization, the model achieves accurate and globally unambiguous phase estimation across the full periodic interval using shallow quantum resources.

Our results show that globally accurate phase estimation depends not only on local sensitivity, but also on the organization of the learned quantum measurement representation and the stability of the decoder inverse mapping. Joint optimization reorganizes the quantum feature embedding and latent manifold into globally distinguishable and well-conditioned structures that support robust estimation across the full periodic domain. Consistent with this interpretation, global estimation performance continues improving even after QFI saturation.

We further show that odd-symmetric activation functions improve latent-manifold conditioning and decoding robustness, acting as effective geometric regularizers for global phase estimation. Together, these findings establish representation geometry and inverse-mapping conditioning as key principles for global quantum metrology, providing a general framework for jointly optimizing quantum encoding and classical inference in programmable quantum sensors.

\begin{appendices}

\section{Training algorithm for VQ-CNNI}
\label{appendix:alg:vq-cnni}

\begin{algorithm}[H]
\caption{End-to-end joint optimization for VQ-CNNI}
\label{alg:vqcnni}
\begin{algorithmic}[1]

\Require
    \State Training phases $\{\phi_i\}_{i=1}^{N}$ sampled from $[-\pi,\pi)$;
    \State Initial parameters $(\boldsymbol{\theta},\boldsymbol{\vartheta},\mathbf{w},\mathbf{b})$; 
    \State Learning rate $\eta$; 
    \State Maximum iterations $T$; 
    \State Patience $P$; 
    \State Minimum iterations $M_{\min}$.
    
\Ensure
     Optimized parameters $(\boldsymbol{\theta}^\ast,\boldsymbol{\vartheta}^\ast,\mathbf{w}^\ast,\mathbf{b}^\ast)$.

\State $(\boldsymbol{\theta}^\ast,\boldsymbol{\vartheta}^\ast,\mathbf{w}^\ast,\mathbf{b}^\ast) \gets (\boldsymbol{\theta},\boldsymbol{\vartheta},\mathbf{w},\mathbf{b})$
\State $\text{best\_loss} \gets \infty$
\State $\text{no\_improve} \gets 0$

\For{$t \gets 1$ \textbf{to} $T$}
    \State $\mathcal{L}_{\text{train}} \gets 0$
    \ForAll{$\phi_i$ in training set}
        \State Prepare variational circuit $U(\boldsymbol{\theta},\boldsymbol{\vartheta})$
        \State Encode phase $\phi_i$
        \State Compute measurement probabilities $p_{\boldsymbol{\theta},\boldsymbol{\vartheta}}(\phi_i)$
        \State Aggregate probabilities into $\mathbf{p}_{\boldsymbol{\theta},\boldsymbol{\vartheta}}(m|\phi_i)$
        \State Predict phase $\tilde{\phi}_i = f_{\mathbf{w},\mathbf{b}}(\mathbf{p}_{\boldsymbol{\theta},\boldsymbol{\vartheta}}(m|\phi_i))$
        \State Accumulate training loss $\mathcal{L}_{\text{train}} \gets \mathcal{L}_{\text{train}} + \frac{1}{N}[1-\cos(\phi_i-\tilde{\phi}_i)]$
    \EndFor
    \State Update $(\boldsymbol{\theta},\boldsymbol{\vartheta},\mathbf{w},\mathbf{b})$ using Adam optimizer with step $\eta$ \Comment{minimize $\mathcal{L}_{\text{train}}$}
    \If{$\mathcal{L}_{\text{train}} < \text{best\_loss}$}
        \State $\text{best\_loss} \gets \mathcal{L}_{\text{train}}$
        \State $(\boldsymbol{\theta}^\ast,\boldsymbol{\vartheta}^\ast,\mathbf{w}^\ast,\mathbf{b}^\ast) \gets (\boldsymbol{\theta},\boldsymbol{\vartheta},\mathbf{w},\mathbf{b})$
        \State $\text{no\_improve} \gets 0$
    \Else
        \State $\text{no\_improve} \gets \text{no\_improve} + 1$
    \EndIf
    \If{$t \bmod K = 0$}
        \State Evaluate the current model on test phases to record SWPE and QFI
    \EndIf
    \If{$t > M_{\min}$ \textbf{and} $\text{no\_improve} \ge P$}
        \State \textbf{break} \Comment{Early stopping}
    \EndIf
\EndFor

\State \Return $(\boldsymbol{\theta}^\ast,\boldsymbol{\vartheta}^\ast,\mathbf{w}^\ast,\mathbf{b}^\ast)$

\end{algorithmic}
\end{algorithm}
\end{appendices}

\bigskip\noindent

\bibliographystyle{unsrt}  
\bibliography{references}
\end{document}